\documentclass[12pt]{article}
\usepackage{graphicx}
\begin{document}
\title{Neural network predictions for $Z'$ boson\\
within LEP2 data set of Bhabha process}
\author{A.N.Buryk, V.V.Skalozub\\
{\small Dnipropetrovs'k National University}\\
{\small Dnipropetrovs'k, 49025 Ukraine}}
\date{01 May, 2009}
\maketitle
\begin{abstract}

The neural network approach is applied to search for the
$Z'$-boson within the LEP2 data set for  $e^+ e^- \to e^+ e^-$
scattering process. In the course of the analysis, the data set is
reduced by 20 percent. The axial-vector $a_e^2$ and  vector
$v_e^2$ couplings of the $Z'$ are estimated at 95\% CL within a
two-parameter fit. The mass is determined  to be $0.53 \le m_{Z'}
\le 1.05 $ TeV. Comparisons with other results are given.
\end{abstract}

\section{Introduction}
A lot of extended models includes a so-called $Z'$ gauge boson -
massive neutral vector particle associated with an extra $U(1)$
subgroup of underlying group \cite{Leike}. In particular, this
particle is predicted by numerous grand unification models. In all
the models, the Abelian $Z'$-boson is described by a low-energy
$\tilde{U}(1)$ gauge subgroup originated in some symmetry breaking
pattern. Searching for $Z'$ is one of the goals of future
experiments at LHC \cite{Rizzo08}, and current ones at Tevatron.
It can manifest itself  either as a real or a virtual state
dependently on the value of its mass.

With this goals keeping in mind, it is reasonable to take into
consideration all accessible nowadays information about
$Z'$-boson, following from different experiments established at
low energies. In particular, in LEP searching for $Z'$ it was
established a model dependent approach and   low bounds on its
mass $ m_{Z'} > 400 $ GeV and/or $m_{Z'} > 800$ GeV  have been
obtained \cite{old_z}, which are dependent  on a specific model.
No actual signals of the $Z'$ have been detected. In recent
experiments at Tevatron the derived low bounds are a little bit
larger and correspond to the masses $m_{Z'} > 850$ GeV
\cite{Rizzo08}.

On the other hand, in series of papers
\cite{Gulov2004}\cite{Gulov2007} a model independent method of
searching for the $Z'$ was developed. This approach accounts for a
renormalizability of an underlying unknown in other respect
theory. The requirement of renormalizability results in a series
of relations between low energy parameters describing interactions
of the $Z'$ with fermions of the standard model. This reduces the
number of low-energy parameters which must be fitted in
experiments.  In this way the couplings of $Z'$ to the fermions of
the standard model and the mass $m_{Z'}$ were estimated at $1-2
 ~\sigma$ CL in the one-parameter \cite{Gulov2004} and many-parameter \cite{Gulov2007} fits. If
 these relations are not taken into consideration, no signals
 (hints, in fact) follow.

It was also concluded that the statistics of the LEP experiments
is not too large to detect the $Z'$ as the virtual state with
enough high precision. So, some further analysis is reasonable.
One needs in the estimates
 which could be used in future experiments. To
determine them in a maximally full way we address to the analysis
based on the predictions of the neural networks (for applications
in high energy physics see, for example, \cite{dudko}). The main
idea of this approach is to constrain a given data set in such a
way that an amount of the data  is considered as an inessential
background and omitted. The remaining data are expected to give a
more precise fit of the parameters of interest. This is the goal
of the present investigation. We treat the full data set on the
Bhabha scattering process obtained in LEP2 experiments by using
the neural network method in order to determine the couplings to
the fermions and the mass of the $Z'$ boson.

\section{Neural network(NN) predictions}

The lack of statistics in LEP experiments does not allow the
determination of the $Z'$-boson mass with CL more than $2\sigma$
in one-parameter fit \cite{Gulov2004} and more than $1\sigma$ in
two-parameter fit \cite{Gulov2007}. We propose two points to
overcome this restriction for the case of the Bhabha scattering
process when a many parameter fit is applied.

First, an increase in parametric space could be compensated by an
increase in the data set, if all possible information is included
in consideration. In this paper we take into consideration the
differential cross-sections   measured by the L3 Collaboration at
183-189 GeV, OPAL Collaboration at 130-207 GeV, ALEPH
Collaboration at 130-183 GeV and the cross-sections obtained by
the DELPHI Collaboration at energies 189-207 GeV. They form the
set of differential cross-sections and their uncertainties for the
center of mass energies.

Second, we use NNs to predict the data set of the investigated
process with increased statistics.

In experiments, the $4\pi$ scattering angle is divided into bins
where the detectors are placed. For each bin we have a
differential cross-section and an uncertainty \cite{data_set}.
 The $Z'$  extends the SM and therefore  contributes  to the differential
cross-section
\begin{equation}
\frac{d\sigma}{dz}=\frac{d\sigma^{SM}}{dz} +
\frac{d\sigma^{Z'}}{dz},
\end{equation}
where $ z = \cos \theta$, $\theta$ is a scattering angle,
$\frac{d\sigma^{SM}}{dz}$ is the contribution coming due to the SM
particles and $\frac{d\sigma^{Z'}}{dz}$ is the contribution due to
the $Z'$ presence. In actual calculations, the first term was
taken from the results reported by the LEP Collaborations and the
second one has been  calculated in the improved Born approximation
with the relations due to renormalizability been taken into
account. In this case it is possible to construct the observables
which uniquely pick out the $Z'$ virtual state
\cite{Gulov2004},\cite{Gulov2007}.

 Hence, the signal of the $Z'$ could be searched in the deviation
of experimental differential cross-section from the SM
differential cross-section.

Taking into account all the noted above we  prepare the needed
experimental data set.  It consists of a differential
cross-section for the Bhabha process and an uncertainty. In order
to use it in our analysis we  subtract from it  the SM
differential cross-section for the Bhabha process. Going this way
we obtain the data set:

\begin{enumerate}
\item scattering sector(bin).
\item $\Delta\frac{d\sigma}{dz}$ - the difference between a
 measured differential cross-section and the corresponding SM differential
 cross-section.
\item $\varepsilon$ - uncertainty.
\end{enumerate}

\begin{figure}[t]
{\includegraphics[width=12cm]{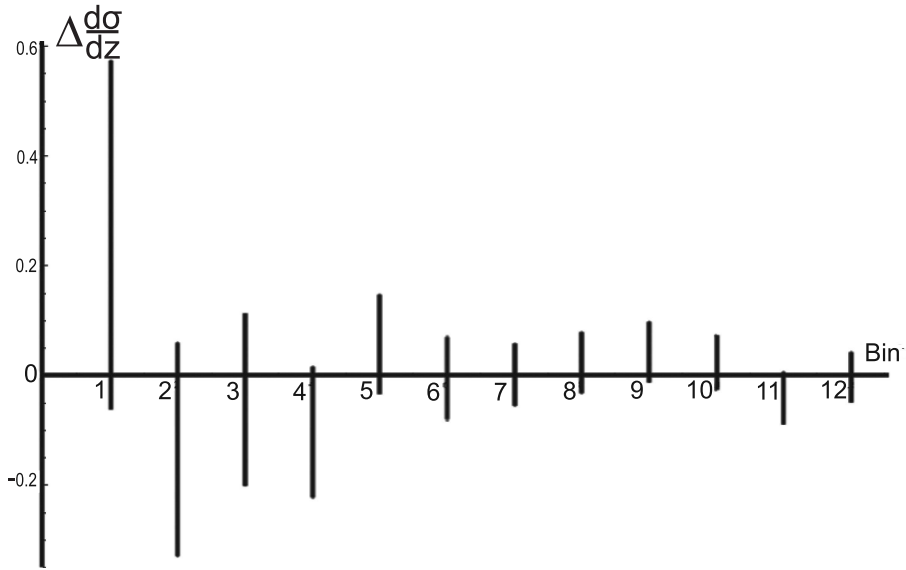}} \caption {Experimental
data are segments  resulting from the experimental uncertainties.
Bins are enumerated. 1-5 bins correspond to scattering in the
backward direction ($Cos\theta<0$), the others - forward direction
($Cos\theta>0$).}\label{fig:fir}
\end{figure}

An overall distribution of the data is shown in Fig.\ref{fig:fir}.
As it is seen, this forms the set of segments determining the
value of the cross-section with its uncertainty. To obtain a more
strict constraint on the axial-vector $\bar{a}^{2}$ and vector
$\bar{v}^{2}$ couplings squared, we have to decrease the lengths
of segments. For this purpose the NN analysis is applied.

At first stage, the NNs were trained to recognize the signal and
the background.

The  contribution of the $Z'$ to a differential cross-section for
the Bhabha process obtained within the SM extended by $Z'$ was
taken  as the signal. Analytic expression for this contribution
reads \cite{Gulov2004}:

\begin{equation}
\frac{d\sigma^{Z'}}{dz}=F_{v}(\sqrt{s},z)\bar{v}^{2}+F_{a}(\sqrt{s},z)\bar{a}^{2}+F_{av}(\sqrt{s},z)\bar{a}\bar{v}
\end{equation}
where $F_{a}$ , $F_{v}, F_{av}$ are the functions depend on the SM
couplings. In actual analysis they have been calculated with
accounting for the relations due to renormalizability. This
contribution was computed for the values of  $0.0 \leq \bar{v}^{2}
\leq 4 \times 10^{-4}$ and $0.0 \leq \bar{a}^{2} \leq 4 \times
10^{-4}$. Such the choice  is based on the results, obtained in
\cite{Gulov2004}.

As the background we set  the deviation from the signal  equaled
to the  redoubled uncertainty of LEP2 experimental data. Thus, the
network was trained to discard data, which correspond to large
deviations from the signal, but include the ones corresponding to
the probable signal of the $Z'$.

The NN processing makes the cutoff of the data. The cutoff
eliminates the data which are assumed  to be the background
signal. Side by side  with the background signals the NN could
discard the signal data too. But total  amount of these events is
negligibly  small as compared to the  amount of discarded
background points.

To create and train NNs the MLPFit program \cite{MLPFit} was used.
Three-layer NNs were used, with 2 neurons in input, 10 neurons in
hidden and 1 neuron in output layer.

An input vector for the networks consists of the scattering sector
and the differential cross-section for this sector. The training
algorithm with back propagation of errors was used. The type of
training - with tutor - was applied. We also worked out  a
necessary computer program to solve the problem.

\begin{figure}[t]
{\includegraphics[width=12cm]{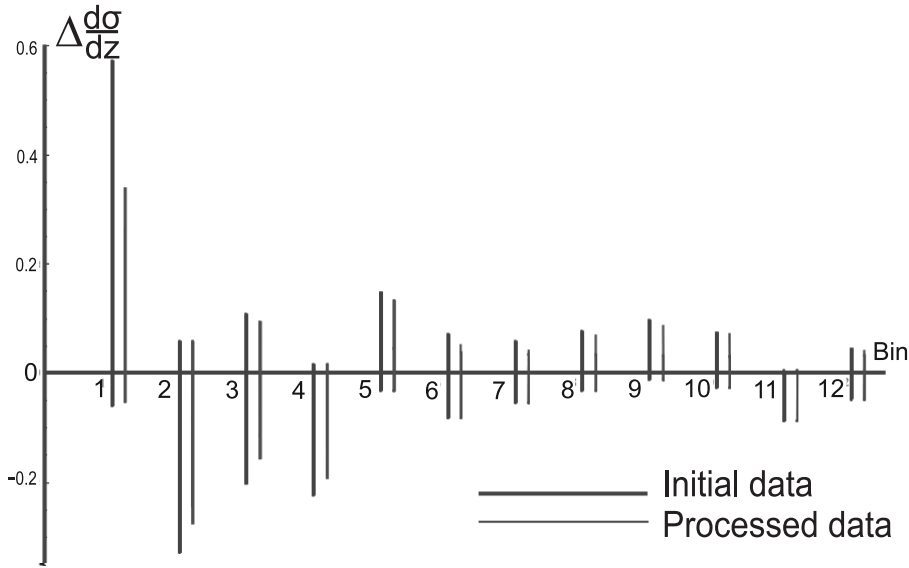}} \caption{Plot of $e^+ e^-
\to e^+ e^-$ process data before and after NN
processing.}\label{fig:sec}
\end{figure}

During processing,  the NNs discarded all the data that have
produced a less than 0.9 at NN output. After NN processing, the
length of segments was decreased.  The total decrease of them is
3-27 percent  (these values vary for different data sets). The
comparative general plot of data set before and after the
application of the NN is shown in Fig.\ref{fig:sec}.

After NNs processing, the data were analyzed by means of the
$\chi^{2}$ method. Denoting an observable by $\sigma_{i}$, one can
construct the $\chi^{2}$-function

\begin{equation}
\chi^{2}(\bar{a},\bar{v}_{e})=\sum\limits_{i}[\frac{\Delta\sigma_{i}^{exp}-\Delta\sigma_{i}^{th}(\bar{a},\bar{v}_{e})}{\delta\sigma_{i}}]^{2},\label{hi}
\end{equation}

where $\Delta\sigma^{exp}$ and $\delta\sigma$ is the deviation of
the measured cross-section from the SM one and the uncertainty for
the Bhabha process, and $\Delta\sigma^{th}$ is the calculated
 $Z'$ contribution.  The sum in Eq.(\ref{hi}) refers to either the data in one
specific process or the combined data for several processes. By
minimizing the $\chi^{2}$-function, the maximal-likelihood
estimate for the $Z'$ couplings was derived. The confidence area
in the parameter space $(\bar{v},\bar{a})$ corresponding to the
probability $\beta$ is defined as

\begin{equation}
\chi^{2} \le \chi^{2}_{min}+\chi^{2}_{CL,\beta}
\end{equation}
where $\chi^{2}_{CL,\beta}$ is the $\beta$-level of the
$\chi^{2}$-distribution with 2 degrees of freedom.

\begin{figure}[t]
{\includegraphics[width=12cm]{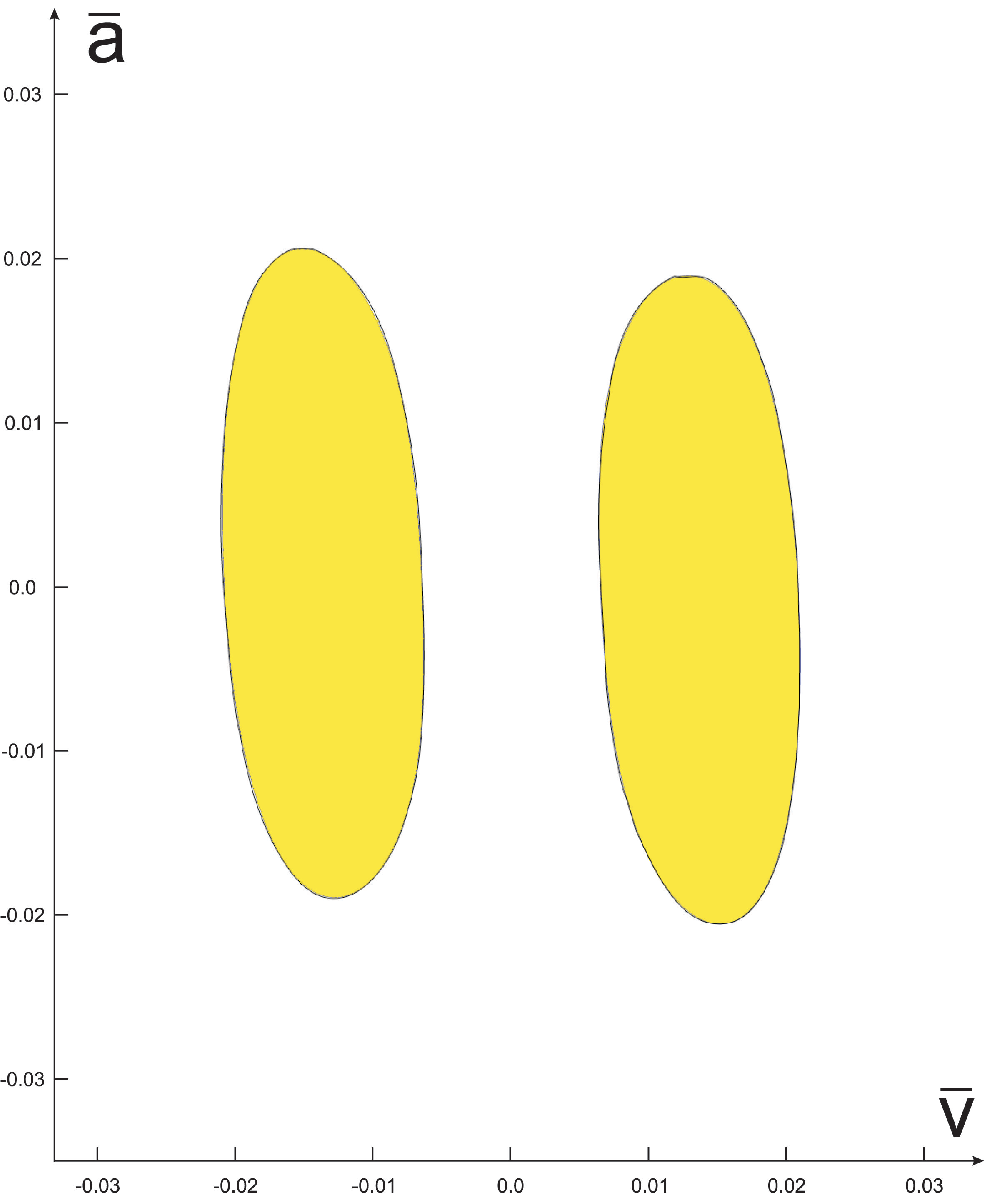}} \caption{The 95\% CL in
$\bar{v}-\bar{a}$ plane for $\bar{v}^{2}$ observable. The data set
after the NN processing.}\label{fig:v2}
\end{figure}

\begin{figure}
{\includegraphics[width=12cm]{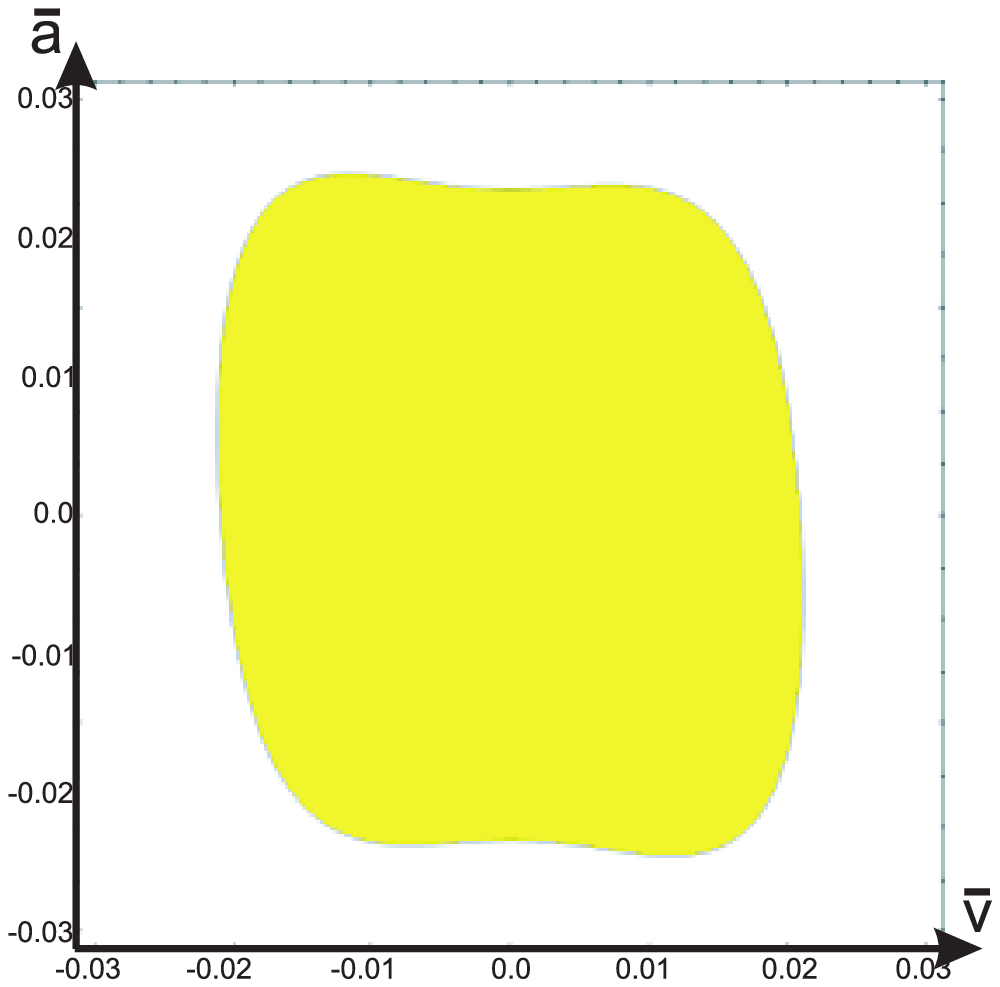}} \caption{The 95\% CL in
$\bar{v}-\bar{a}$ plane for $\bar{v}^{2}$ observable. The data set
before the NN processing.}\label{fig:v2old}
\end{figure}

The analysis of the $\chi^{2}$-function for the $\bar{v}^{2}$
observable gives  $\chi^{2}_{min}=158.49$ at $\bar{v}^{2} = 2.37
\times 10^{-4}$. The 95\% CL area ($\chi^{2}_{CL}=5.99$) is shown
in Fig.\ref{fig:v2}. From this area the value $\bar{v}^{2} =
(2.4\pm 1.99)\times 10^{-4}$ is obtained. The 95\% CL area for
data before the NN processing is shown in Fig.\ref{fig:v2old}.

\begin{figure}[t]
{\includegraphics[width=12cm]{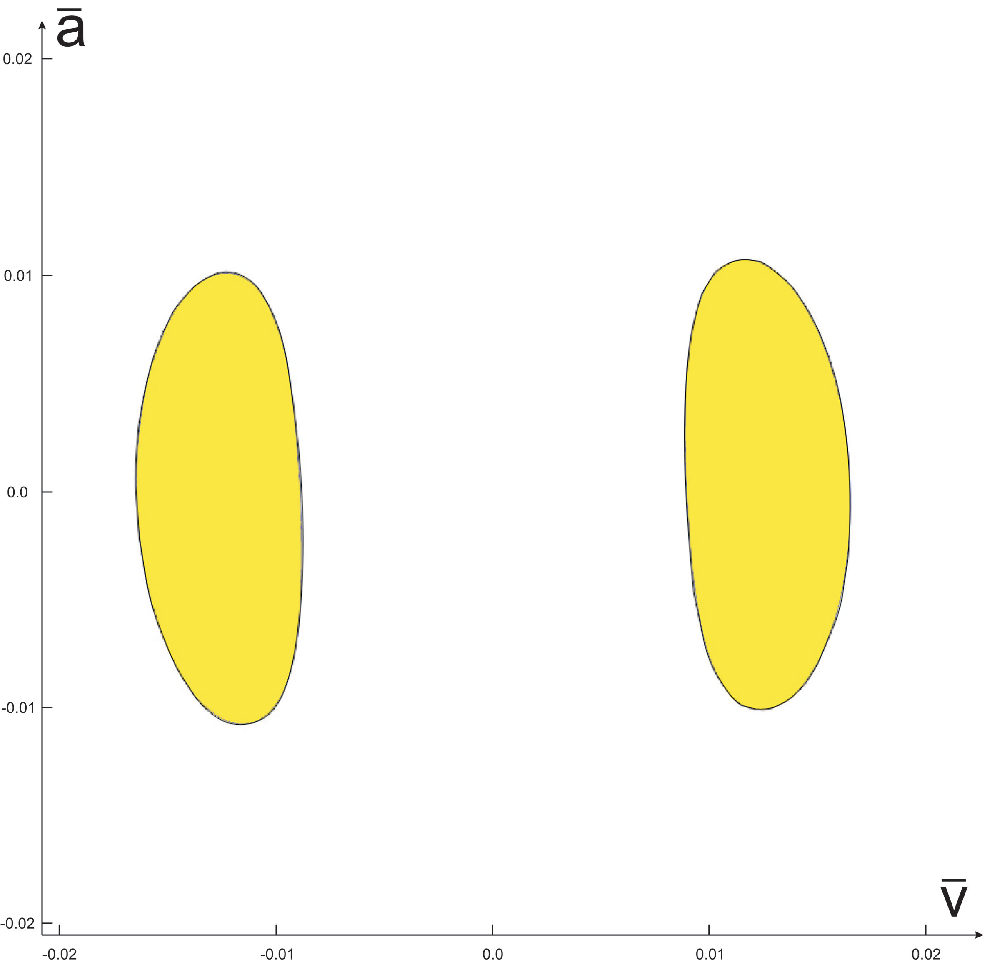}} \caption{The 95\% CL in
$\bar{v}-\bar{a}$ plane for $\bar{a}^{2}$ observable. The data set
after the NN processing. As it is seen, the zero point of
$\bar{a}$ is inside the confidence area.}\label{fig:a2}
\end{figure}

\begin{figure}
{\includegraphics[width=12cm]{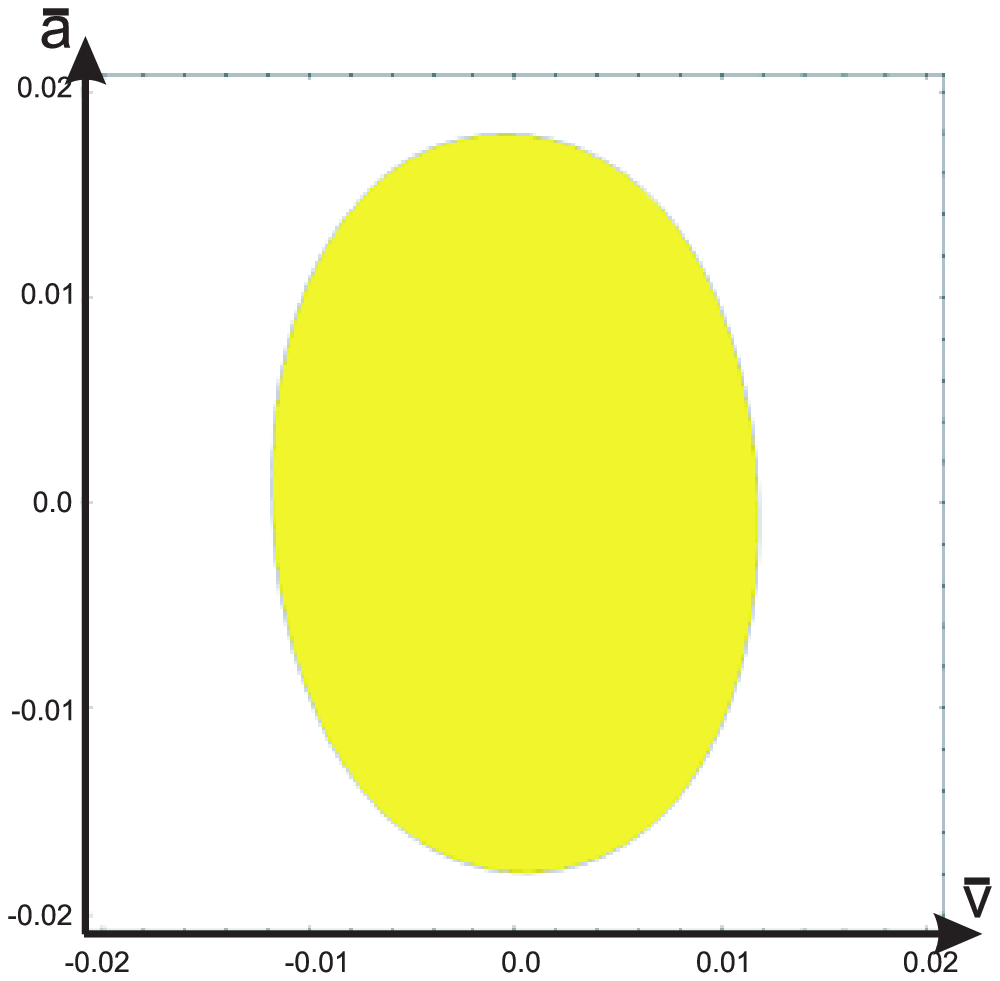}} \caption{The 95\% CL in
$\bar{v}-\bar{a}$ plane for $\bar{a}^{2}$ observable. The data set
before the NN processing.}\label{fig:a2old}
\end{figure}

The analysis of the $\chi^{2}$-function for the $\bar{a}^{2}$
observable results in  $\chi^{2}_{min}=629.66$ at $\bar{a}^{2}
=1.6 \times 10^{-7}$. The 95\% CL area ($\chi^{2}_{CL}=5.99$) is
shown in Fig.\ref{fig:a2}. To compare the results,  the  95\% CL
area for data before NN processing is shown in
Fig.\ref{fig:a2old}. As one can see, the zero point of $\bar{a}$
is inside the confidence area. So, we are able to obtain only an
upper limit on its value $\bar{a}^{2} \le 1.1\times 10^{-4}$ at
the $2\sigma$ CL.

To constrain the value of $Z'$ mass by the derived bounds on the
four-fermion couplings $\bar{v}^{2}$ let us assume that the
coupling $\bar{g}$ is of the order of the SM gauge couplings,
$\bar{g}^{2}/(4\pi) \simeq 0.01-0.03$. Then the obtained value
$\bar{v}^{2}$ corresponds to $m_{Z'}=0.53-1.05$ TeV.

In Ref. \cite{Gulov2004} the value $\bar{v}^{2} = 2.18 \pm 1.82
\times 10^{-4}$ was obtained for the Bhabha process at the
$2\sigma$ CL within one-parameter fit.

\begin{table}
\begin{center}
\caption{Constraints on the $Z'$ four-fermion couplings
$\bar{v}^{2}$ obtained in different investigations}
\begin{tabular}{|c|c|c|c|}
 \hline
\multicolumn{2}{|c|}{one-parameter fit} &
two-parameter fit & NN two-parameter fit\\
\hline $1\sigma$ & $2\sigma$ & $2\sigma$ &
$2\sigma$\\
\hline \multicolumn{4}{|c|}{$\times 10^{-4}$}\\
\hline $2.24\pm 0.92$ & $2.18 \pm 1.82$ & $\le 1.44$ &
$2.4\pm 1.99$\\
 \hline
\end{tabular}
\end{center}
\end{table}

\section{Discussion}

It is shown in Fig.\ref{fig:sec} how the NN selects  data
corresponding to the background. We note once again that the  NN
discards not only a background but also the signal. Nevertheless,
the discarded signal data are negligibly small as compared to the
discarded background ones.

The obtained values of $\bar{v}^{2} = (2.4\pm 1.99)\times 10^{-4}$
and $\bar{a}^{2} \le 1.1\times 10^{-4}$ are comparable to the
results obtained in \cite{Gulov2004}. In this paper the results
were obtained within the one-parameter fit. In paper
\cite{Gulov2007} it was shown that there is only a $1\sigma$ CL
hint in the two-parameter fit. The analysis carried out in the
present investigation  shows the $2\sigma$ CL hint of the $Z'$.
The estimate of the mass value gives $m_{Z'}=0.53-1.05$ TeV. This
is signaling a comparably  not heavy $Z'$, that is of interest for
the  experiments at the Tevatron and LHC. The estimated couplings
are also important to analyze the current and future experimental
data.

To conclude, we note that the derived predictions of the NNs
analysis of the two parameter fits are in good agreement with
other one-parameter model-independent fits accounting for the $Z'$
gauge boson existence \cite{Ferroglia}, \cite{Gulov2007}.

\end{document}